\documentstyle[12pt,preprint]{aastex}
\newcommand{\bn}{{\mathbf{\nabla}}}
\shorttitle{On the Physical Origin of the Fundamental Plane}
\shortauthors{O\~norbe et al.}
\begin{document}
 
\title{ Clues on the Physical Origin of the Fundamental Plane from Self-consistent
Hydrodynamical Simulations}
 
\author{J. O\~norbe
\footnote{Dpt.\ F\'{\i}sica Te\'orica C-XI,
Universidad Aut\'onoma de Madrid,
E-28049 Cantoblanco,
Madrid,
Spain;
 $^2$ Current address:
 Dept.\ of Physics, Mahidol University, Bangkok 10400, Thailand; $^3$ Dpt.\ F\'{\i}sica y A.C., Universidad Miguel Hern\'andez,
 E-03206 Elche, Alicante, Spain},
 R. Dom\'{\i}nguez-Tenreiro$^{1}$,
 A. S\'aiz$^{1, 2}$,
  A. Serna$^{3}$, H. Artal$^{1}$
  }
\email{
jose.onnorbe@uam.es, rosa.dominguez@uam.es, alex@astro.phys.sc.chula.ac.th,
arturo.serna@umh.es, hector.artal@uam.es}

\begin{abstract}    
We report on a study of the parameters characterizing the mass and velocity
distributions of two samples of relaxed elliptical-like-objects (ELOs)
identified, at $z=0$, in a set of self-consistent hydrodynamical simulations
operating in the context of a concordance cosmological model.
Star formation (SF) has been implemented in the simulations
in the framework of the turbulent sequential scenario through a 
phenomenological parameterization that takes into account
stellar physics processes implicitly through the values
of a threshold gas density
and an efficiency parameter.
Each ELO sample is characterized by the values these parameters take. 
We have found that the (logarithms of the)
ELO stellar masses,
projected half- stellar mass radii, and stellar central
l.o.s. velocity dispersions define 
{\it dynamical} Fundamental Planes (FPs).
Zero-points depend on the particular values that 
the SF parameters take, while  slopes do not change.
The ELO samples have been found to show systematic trends with the mass
scale in both, the relative content and the relative distributions of the
baryonic and the dark mass ELO components. The physical origin of
these trends lies in the systematic decrease, with increasing ELO mass, 
of the relative 
dissipation experienced by the baryonic mass component along ELO
mass assembly,  resulting into a tilt of the dynamical FP
relative to the virial plane. 
ELOs also show kinematical segregation,
but it does not appreciably change
with the mass scale.
We have found that the dynamical FPs shown by the two ELO samples
are consistent with that shown by the SDSS elliptical sample in the 
same variables, with
 no further need for any relevant contribution
from stellar population effects to explain the observed tilt.
These effects could, however, have contributed to the scatter of the
observed FP, as the dynamical FPs have been found to be thinner
than the observed one.
The results we report on hint, for the first time, to a possible way
to understand the tilt of the observed FP in a 
cosmological context.

\end{abstract}
 
 
\section{INTRODUCTION}  

The 3-parameter space of the observed effective radius, $R_e^{\rm light}$,
the mean surface brightness within that radius, $<I^{\rm light}>_e$,
and the central line-of-sight (los) velocity dispersion,
$\sigma_{\rm los, 0}$, of early-type galaxies is not homogeneously
populated. These galaxies define a plane on this space, known as
the Fundamental Plane (FP, Djorgovski \& Davis 1987; Dressler et 
al. 1987a; Faber et al. 1987; Kormendy \& Djorgovski 1989) defined by:

\begin{equation}
\log_{10} R_e^{\rm light} = a \log_{10} \sigma_{\rm los, 0} + b \log_{10} <I^{\rm light}>_e + c.
\end{equation}

A new standard of
reference for nearby elliptical  galaxies is provided by
the Sloan digital sky survey (SDSS, see York et al. 2000)
sample of early-type galaxies (see Bernardi et al. 2003a, 2003b, 2003c
for the Early Data Release),
containing to date 9000 morphologically selected ellipticals
from different environments,
a number larger than  the number of ellipticals in all the previously
analyzed samples.
The values of the FP coefficients from this sample are $a \simeq 1.5$,
similar in the four SDSS bands, 
$b \simeq -0.77$, and $c \simeq -8.7$ (see their exact values in   Bernardi et al. 2003c,
Table 2) with a small scatter.
These SDSS results confirm previous ones,
either in the
optical (Lucey, Bower \& Ellis 1991; de Carvalho \& Djorgovski 1992; 
Bender, Burstein \& Faber 1992; Jorgensen et al. 1993;
Prugniel \& Simien 1996;  
Jorgensen et al. 1996)
or in the near-IR wavelenghts (Recillas-Cruz et al. 1990, 1991;
Pahre, Djorgovski \& de Carvalho 1995;
Mobasher et al. 1999), even if the published values of $a$ show larger
values in the $K$-band
than  at shorter wavelengths 
(see, for example, Pahre, de Carvalho \& Djorgovski 
1998).

The existence of the FP and its small scatter has the important
implication that it provides us with a strong constraint 
when studying elliptical galaxy formation and evolution
(Bender, Burstein \& Faber 1993; Guzm\'an, Lucey \& Bower 1993;
Renzini \& Ciotti 1993).
The physical origin of the FP is not yet clear, 
but it must be a consequence of the physical processes
responsible for galaxy assembly.
These processes built up early type galaxies as dynamically hot systems
whose configuration in phase space are close to equilibrium.
Taking an elliptical galaxy as a system in equilibrium, 
a mass scale $M_{\rm FP}$
can be estimated from the observable $R_e^{\rm light}$,
$<I^{\rm light}>_e$
and $\sigma_{\rm los, 0}$   parameters through the expression:
\begin{equation} 
M_{\rm FP} = 3 c_{\rm M} \sigma_{\rm los, 0}^{2} R_e^{\rm light} / G,
\label{MFP}
\end{equation}
where $G$ is the gravitational constant and $c_{\rm M} $ is 
a mass structure coefficient.
The total luminosity of the galaxy is given by the expression
$L = 2 \pi <I^{\rm light}>_{e} (R_e^{\rm light})^{2}$.
As Faber et al. (1987) first pointed out, the virial theorem
$M_{\rm vir} = c_{\rm F} (\sigma_{3}^{\rm tot})^{2} r_{e}^{\rm tot}/ G $,
with $\sigma_{3}^{\rm tot}$ the average 3-dimensional velocity dispersion
of the whole elliptical, including both dark and baryonic matter,
$r_{e}^{\rm tot}$ the dynamical half-radius or radius enclosing 
half the total mass of the system, and $c_{\rm F}$ a form factor
of order unity, would imply
\begin{equation} 
R_e^{\rm light} = {3 c_{\rm M}^{\rm vir} L \sigma_{\rm los, 0}^{2} <I^{\rm light}>_{e}^{-1} \over 2 \pi G M_{\rm vir}}
\label{virrel}
\end{equation}
where $c_{\rm M}^{\rm vir}$ is defined by Eq. (\ref{MFP}) when
$M_{\rm FP} = M_{\rm vir}$,
that is
\begin{equation}
c_{\rm M}^{\rm vir} = {G M_{\rm vir} \over 3 \sigma_{\rm los, 0}^{2} R_e^{\rm light}},
\end{equation}
or
\begin{equation}
c_{\rm M}^{\rm vir} = c_{\rm F} c_{\rm v} c_{\rm r}
\end{equation}
with
$c_{\rm v} = (\sigma_{3}^{\rm tot})^{2}/3 \sigma_{\rm los, 0}^{2}$ and
$c_{\rm r} = r_{e}^{\rm tot}/R_e^{\rm light}$.
Assuming that both the quantities  $c_{\rm M}^{\rm vir}$ and
the dynamical mass-to-light ratios, $M_{\rm vir}/L$, are independent of mass,
the scaling relation 
$R_e^{\rm light} \propto \sigma_{\rm los, 0}^{2} <I^{\rm light}>_{e}^{-1}$  
would then hold. However, this predicted scaling law is inconsistent
with those found observationally ($a \neq 2, b \neq -1$), i.e.,
the FP is tilted relative to the virial relation,
implying that at least some of the assumptions
made to derive it is incorrect. 

Different authors interpret the tilt of the FP relative to the 
virial relation as caused by different misassumptions that we comment
briefly (note that we can write $M_{\rm vir}/L =  M^{\rm star}/L \times M_{\rm vir}/M^{\rm star}$,
where $M^{\rm star}$ is the stellar mass of the elliptical galaxy):
i) A first possibility
is that the tilt is due to
systematic changes of stellar age and metallicity with galaxy mass, or, even,
to changes of the slope of the stellar initial mass function (hereafter, IMF)
with galaxy mass, resulting in systematic changes in the
{\it stellar}-mass-to-light ratios, $M^{\rm star}/L$, with mass
or luminosity
 (Zepf \& Silk 1996; Pahre at al. 1998; Mobasher et al. 1999).
But these effects  could explain at most only 
$\sim$ one third of the $\beta \neq 0$ value 
in the $B$-band (Tinsley 1978; Dressler et al. 1987;
 Prugniel \& Simien 1996; see also 
Renzini \& Ciotti 1993;
Trujillo, Burkert \& Bell 2004).
Futhermore,  early-type galaxies in the SDSS have been found to have
roughly constant stellar-mass-to-light ratios (Kauffmann et al. 2003a, 2003b).
Anyhow, the presence of a tilt in the $K$-band FP, 
where population effects are no important, indicates
that it is very difficult that the tilt
is caused by stellar physics processes alone,
as Bender et al. (1992), Renzini \& Ciotti (1993),
Guzm\'an et al. (1993),
Pahre et al. (1998), 
among other authors, have suggested.
ii) A second possibility is that $M_{\rm vir}/L$ changes systematically 
with the mass scale because the total dark-to-visible mass ratio,
$M_{\rm vir}/M^{\rm star}$ changes 
(see, for example, 
Renzini \& Ciotti 1993; Pahre et al. 1998;
Ciotti, Lanzoni \& Renzini 1996; Padmanabhan et al. 2004).
Otherwise, a dependence of $c_{\rm M}^{\rm vir}$ on the mass
scale could be caused by systematic differences in
iii), the dark  versus bright matter spatial distribution,
%including homology breaking in the galaxy luminosity profiles,
which could be measured through systematic variations of 
the $c_{\rm r}$ coefficients with mass,
iv), the kinematical segregation,
the rotational  support and/or velocity
dispersion anisotropy in the stellar
component  (dynamical non-homology), measurable through 
the $c_{\rm V}$ coefficients, and,
v), systematic   geometrical effects, measurable through 
the $c_{\rm F}$, $c_{\rm r}$ or $c_{\rm V}$ coefficients.
Taking into account these effects in the FP tilt demands
modellizing the galaxy mass and velocity three-dimensional distributions
and comparing the outputs with high quality data.
Bender et al. (1992) considered  effects iii) and iv);
Ciotti et al. (1996) explore ii) - iv) and conclude that
an systematic increase in the dark matter content with mass, or differences
in its distribution, as well as a dependence of the S\'ersic (1968)
shape parameter for the luminosity profiles 
with mass, may by themselves formally
produce the tilt; Padmanabhan et al. (2004) find evidence of
effect ii) in SDSS data.
Other authors have also shown that allowing for broken homology,
either dynamical (Busarello et al. 1997), in the luminosity
profiles (Trujillo et al. 2004), or both 
(Prugniel \& Simien 1997;
Graham \& Colless 1997; Pahre et al. 1998),
brings the observed FP closer to the virial plane.

One important source of ambiguity in observational data analysis
comes from the impossibility
to get accurate measurements of the elliptical
three-dimensional  mass distributions
(either dark, stellar or gaseous) and velocity distributions.
Analytical models give very interesting insights into
these distributions as well as the physical
processes causing them, but are somewhat
limited by symmetry  considerations and other necessary  simplifying
hypotheses.
Self-consistent gravo-hydrodynamical 
simulations are  a very convenient tool to work out this problem,
as they {\it directly} 
provide  with complete 6-dimensional
phase-space information on each constituent particle sampling a 
given galaxy-like object formed in the simulation,
that is, they give directly the mass and velocity  distributions
of dark matter, gas and stars of each objet.  
This  phase space information allows us
 to test whether or not 
  the $c_{\rm M}^{\rm vir}$ (that is, the $c_{\rm F},
  c_{\rm v}$ and $c_{\rm r}$) coefficients,
  as well as the $M_{\rm vir}/M^{\rm star}$ ratios, do or do not 
  systematically depend on the mass scale.
  This is the issue addressed in this Letter,
  where we analyze whether the dependence is such that
  the tilt and the scatter of the observed FP can be explained
in terms of the  regularities in the structural and dynamical
properties of ELOs formed in self-consistent
hydrodynamical simulations.

\section{The Dynamical Fundamental Plane of Simulated Objects}

In a self-consistent numerical approach
 initial conditions are set at high $z$ as a Montecarlo
realization of the field of primordial fluctuations
in a given cosmological model; then the evolution 
of these fluctuations is numerically followed
up to $z =0$ by means of a computing code that solves the   N-body plus hydrodynamical
evolution equations.
We have run ten simulations in the framework of
a flat $\Lambda$CDM cosmological model, with $\Omega_{\Lambda}=0.65$,
 $\Omega_{\rm baryon}=0.06$, $\sigma_8=1.18$ and $h=0.65$. 
The code used in our simulations is DEVA
(Serna, Dom\'{\i}nguez-Tenreiro, \& S\'aiz 2003). In this code, particular
  attention has been paid to the implementation of 
conservation laws (energy; entropy, taking into account the $\bn h$
terms; angular momentum).
Star formation (SF) has  been implemented in the code
in the framework of the turbulent sequential scenario
(Elmegreen 2002) through a phenomenological
parameterization,
  that transforms   cold locally-collapsing gas,
  denser than a threshold density,
  $\rho_{\rm thres} $,
  into stars with a timescale given
  by the empirical Kennicutt-Schmidt law (Kennicutt 1998),
  with an average star formation efficiency at the scales
  resolved by the code $c_{\ast}$;
  possible feedback effects are implicitely taken into account
  through the values of the SF parameters.
  In any run, 64$^3$ dark matter and 64$^3$ baryon  particles,
  with a mass of $1.29 \times
  10^8$ and $2.67 \times 10^7 $M$_{\odot}$, respectively, have been used
to  homogeneously sample the density field in a periodic box of 10 Mpc side. 
  We refer the reader to Serna et al. 2003
    and to S\'aiz, Dom\'{\i}nguez-Tenreiro, \& Serna 2004 for further details
      on the simulation technique and SF implementation in the code.
Five out of the ten simulations 
(the SF-A type simulations)
share the SF parameters ($\rho_{\rm thres} = 6 \times 10^{-25} $ gr cm$^{-3}$,
$c_*$ = 0.3) and differ in the seed  used to build up the initial conditions.
To test the role of SF parameterization,
the same initial conditions
have been run with different SF parameters
($\rho_{\rm thres}$ =  $1.8 \times 10^{-24} $ gr cm$^{-3}$,
$c_*$ = 0.1) making SF  more difficult, contributing another set of five
simulations  (hereafter, the SF-B type simulations). 
Galaxy-like objects of different morphologies appear in the
 simulations.   
 ELOs have been identified as those objects
 having a prominent
         dynamically relaxed stellar spheroidal component,
with no  disks and very low cold gas
	content. This stellar component
	has typical sizes of no more than $\sim $ 10 - 40  kpc and it
	is embedded in a halo of dark matter typically ten  times larger in size.
	ELOs have  also an extended corona of hot diffuse gas.
It turns out that  26 (17) out of the 
more massive objects formed in SF-A (SF-B) 
type simulations fulfil this condition,
giving the hereafter termed  SF-A and SF-B ELO samples. 
Note that due to their respective  SF implementations,
galaxy-like objects formed in
SF-A type simulations tend to be of earlier type than their counterparts 
formed in SF-B type simulations; this is why the ELO number in SF-A sample
in higher than in SF-B sample. Moreover, gas has had more time to lose
energy along SF-B type ELO assembly  than in their SF-A type counterparts,
and, consequently, the former have smaller sizes than the latter.

Following the discussion in $\S$1,
two mass scales have been measured on  ELOs:
 the virial mass $M_{\rm vir}$, or total mass of the ELO at its halo scale
 (we adopt the fitting formula of Bryan \& Norman, 1998, for the 
 spherical overdensity at virialization), and
 the stellar mass at the ELO scale, $M_{\rm bo}^{\rm star}$.
 Concerning length scales, the relevant ones are: i), at the halo
 scale: the virial radii, $r_{\rm vir}$ and the half-total mass radii,
 $r_{\rm e, h}^{\rm tot}$, enclosing $M_{\rm vir}/2$; 
 and ii), at the ELO scale,
 the stellar  half-mass radii, $r_{\rm e, bo}^{\rm star}$,
     defined as those radii enclosing half the $M_{\rm bo}^{\rm star}$ mass,
     and the  projected stellar  half-mass radii, $R_{\rm e, bo}^{\rm star}$,
 measured onto the projected
mass distribution.
The total (including baryons and dark matter)
mean square velocity within $r_{\rm vir}$, $\sigma^{\rm tot}_{\rm 3, h}$,  
as well as  the  mean square stellar and
central stellar\footnote{Recall that the empirical
 l.o.s velocity dispersion, 
$\sigma_{\rm los, 0}$, is measured through {\it stellar spectra}}
l.o.s velocity dispersions, $\sigma_{\rm 3, bo}^{\rm star}$ and
$\sigma^{\rm star}_{\rm los, 0}$,
respectively, have also  been measured on ELOs. 
Note that the scales entering  the virial relation are
$r_{\rm e, h}^{\rm tot}$ and
$\sigma^{\rm tot}_{\rm 3, h}$ and that they are not observationally  available.
Assuming that the projected stellar {\it mass} distribution,
$\Sigma_{\rm star}(R)$, can be taken as a measure of the
surface {\it brightness} profile, then
$<\Sigma_{\rm star}>_{\rm e} = c <I^{\rm light}>_e$, with $c$ a constant, and
$R_{\rm e, bo}^{\rm star} \simeq R_{\rm e}^{\rm light}$
and we can look for a fundamental plane
(hereafter, the dynamical FP) 
in the 3-space of the structural and dynamical parameters
$R_{\rm e, bo}^{\rm star}$, $<\Sigma_{\rm star}>_{\rm e}$ and
$\sigma^{\rm star}_{\rm los, 0}$,
directly provided by the hydrodynamical simulations.   
To make this analysis as clear as possible, we transform to
a  $\kappa$-like orthogonal coordinate system,
the dynamical $\kappa_i^{\rm D}$ system, $i$=1,2,3, 
 similar to that introduced
by Bender, Burstein \& Faber (1992), but 
using $R_{\rm e, bo}^{\rm star}$ instead of $R_{\rm e}^{\rm light}$ and
$<\Sigma_{\rm star}>_{\rm e}$ instead of $<I^{\rm light}>_e$,
and, consequently, free of age,  metallicity  or IMF effects.
The $\kappa^{\rm D}$ and $\kappa$ coordinates are related by the expressions:
$\kappa_1 \simeq \kappa^{\rm D}_1$, 
$\kappa_2 \simeq \kappa^{\rm D}_2 - \sqrt 6/3 \log M_{\rm bo}^{\rm star}/L$
and $\kappa_3 \simeq \kappa^{\rm D}_3  +  \sqrt 3/3 \log M_{\rm bo}^{\rm star}/L$.
We discuss the tilt and the scatter of the dynamical FP  separately.
We first address the tilt issue. We use at this
 stage for the $R_{\rm e, bo}^{\rm star}$ and $\sigma^{\rm star}_{\rm los, 0}$
  variables the averages over three  orthogonal l.o.s.  projections,
to minimize the scatter in the plots caused by projection
  effects.
	   
Figure 1 plots the  $\kappa_{3}^{\rm D}$  versus $\kappa_{1}^{\rm D}$
(top) and $\kappa_{2}^{\rm D}$  versus $\kappa_{1}^{\rm D}$ (bottom)
diagrams for ELOs in the SF-A (filled symbols) and SF-B (open symbols) samples.
We also drew the 2$\sigma$
concentration ellipses in the respective
variables, as well as its major and minor axes,  for
the SDSS early-type galaxy  sample in the $z$ band  as analyzed by
Bernardi et al. 2003b, 2003c\footnote{
The constant stellar-mass-to-light ratios  allow us to write the covariance
matrix using  the $E \equiv \log M_{\rm bo}^{\rm star}$ variable
instead of absolute magnitude or the logarithm of the luminosity $L$}.
We recall that the ellipse major axis corresponds to  the orthogonal mean square
regression line
for the two variables in the Figure 
(for further details see S\'aiz et al. 2004).
  The most outstanding feature of Figure 1 (upper panel) is the good scaling behaviour of
      $\kappa_{3}^{\rm D}$  versus 
      $\kappa_{1}^{\rm D}$,
      with a very low
            scatter (see the slopes $M_1$ in Table 1;
	    note that the slopes for the SF-A and SF-B samples are
consistent within their errors while the zero-points depend on the
SF parameterization through the ELO sizes).
The values of the slopes in Table 1
mean that systematic variations of the structural
 and dynamical properties of ELOs with the mass scale cause, by themselves,
a tilt of the dynamical FP relative to the virial relation.
Another interesting feature of Figure 1
is that it shows that most of the  values of the $\kappa_{i}^{\rm D}$
coefficients are whithin the 2$\sigma$ concentration ellipses in both plots for
ELOs formed in SF-A type simulations, with a slightly worse  agreement
for ELOs in the SF-B sample. This means that ELOs have
counterparts in the real world (S\'aiz et al. 2004).
It is also worth mentioning that these results  are stable against
slight changes in the values of the $\Omega_{\Lambda}$, $\Omega_{\rm baryon}$
or $h$ parameters;
for example, we have tested that 
using their preferred WMAP values shows results negligibly different to
those plotted in  Figure 1.

  To deepen into the causes of the tilt, 
      we have calculated the slope $\beta_{\rm vir}$ of the 
      $ M_{\rm vir}/M_{\rm bo}^{\rm star} \propto  (M_{\rm bo}^{\rm star})^{\beta_{\rm vir}}$  
scaling relation for ELOs in the SF-A and SF-B samples (Table 1).
We got  $\beta_{\rm vir} > 0$,
indicating that the mass fraction of stars
bound to the ELOs (or, more generically, cold baryons)
relative to total mass within $r_{\rm vir}$
decreases with the mass scale 
(as suggested by Renzini \& Ciotti 1993 and Pahre et al. 1998).
We have also found that the $c_{\rm M}^{\rm vir}$ coefficients show
a mass dependence that can be parametrized as a scaling relation
$c_{\rm M}^{\rm vir} \propto  (M_{\rm bo}^{\rm star})^{\beta_{\rm M}}$
 (broken homology, see Table 1).
As discussed in $\S$1, different possible  sources for
homology breaking exist. To dilucidate which of them are relevant,
we first note that the $c_{\rm V}$ and $c_{\rm r}$
coefficients can be written as:
$c_{\rm V} = c_{\rm VD} \times c_{\rm VPC}$,
with
$c_{\rm VD} \equiv (\sigma_{\rm 3, h}^{\rm tot} / \sigma_{\rm 3, bo}^{\rm star})^2$
and 
$c_{\rm VPC} \equiv (\sigma_{\rm 3, bo}^{\rm star})^2 / 3 (\sigma_{\rm los, 0}^{\rm star})^2$
and $c_{\rm r} = c_{\rm rD} \times c_{\rm rP}$ 
with $c_{\rm rD} \equiv r_{\rm e, h}^{\rm tot} / r_{\rm e, bo}^{\rm star}$
and $c_{\rm rP} \equiv r_{\rm e, h}^{\rm star} / R_{\rm e, bo}^{\rm star}$.
This gives $c_{\rm M}^{\rm vir} = c_{\rm F} c_{\rm VD} c_{\rm VPC} c_{\rm rD} c_{\rm rP}$, where the $c_{\rm VD}$ and $c_{\rm rD}$ factors measure
the kinematical and spatial segregations between dark and stellar matter,
respectively, while $c_{\rm VPC}$ and $c_{\rm rP}$ measure
projection and other geometrical effects in the stellar mass and
velocity distributions. Writting
$c_{\rm i} \propto (M_{\rm bo}^{\rm star})^{\beta_{\rm i}}$,
with i = F, VD, VPC, rD, rP, we get 
$\beta_{\rm M} = \beta_{\rm F} + \beta_{\rm VD} + \beta_{\rm VPC} + \beta_{\rm rD} + \beta_{\rm rP}$ when the $\beta_{\rm i}$ slopes are calculated
through direct fits.
These $\beta_{\rm i}$ slopes are given in Table 1, as well as their
95 \% confidence intervals both for the SF-A and SF-B samples.
We see that, irrespective of the SF parameterization,
the main contribution to the homology breaking comes from
the $c_{\rm rD}$ coefficients (see Guzm\'an et al. 1993),
while $\beta_{\rm F}$ and 
$\beta_{\rm VD}$ have values consistent with $c_{\rm F}$ and
$c_{\rm VD}$ being independent of the ELO mass scale,
$c_{\rm rP}$  
and $c_{\rm VPC}$ show a very mild mass dependence in the SF-A
sample and none in the SF-B sample.

We now turn to consider the scatter of the dynamical FP for the ELO samples
and compare it with the scatter of the FP for the SDSS elliptical sample,
calculated as the square root of the smallest eigenvalue of the 3$\times$3
covariance matrix in the $E$ (or $\log L$), 
$V$ and $R$ variables (Saglia et al. 2001).
As Figure 1 (top) suggests, 
when projection effects are circunvented by taking averages
over different directions, the resulting three dimensional orthogonal scatter 
for ELOs is smaller than for SDSS ellipticals ($\sigma_{\rm EVR} = 0.0164$
and $\sigma_{\rm EVR} = 0.0167$ for the SF-A and SF-B samples,
respectively, to be compared with
$\sigma_{\rm LVR} = 0.0489$  for the SDSS
in the $\log L, V $ and $R$ variables).
To estimate the contribution of projection
effects to the observed scatter, we have calculated the  orthogonal
scatter for ELOs when no averages over projection directions 
for the $R_{\rm e, bo}^{\rm star}$ and $\sigma_{\rm los, 0}^{\rm star}$
variables are made. The scatter ($\sigma_{\rm EVR} = 0.0238 $ 
and $\sigma_{\rm EVR} = 0.0214$ for the SF-A and SF-B samples) 
increases, but it is still 
lower than observed. This indicates  that a contribution
from stellar population effects is needed to explain 
 the scatter of the observed FP, as suggested by  different authors 
(see, for example,  Pahre et al. 1998; Trujillo et al. 2004).

To sum up, the ELO samples have been found to show systematic trends
with the mass
scale in both, the relative content and the relative distributions of the
baryonic and the dark mass ELO components.
These trends do not significantly depend on the star formation
parameterization and they are due to
a systematic decrease, with increasing ELO mass, of the relative amount
of  dissipation experienced by the baryonic mass component along ELO
formation, a possibility that
Bender et al. (1992), Guzm\'an et al. (1993), and
Ciotti et al. (1996) had suggested.
These trends cause a tilt of the virial plane 
in such a way that there is no further need of any relevant contribution
from stellar population effects to explain the observed tilt.
The scatter of the observed FP, however,  probably 
requires a contribution from such
stellar effects.

This work was partially supported by the MCyT and MEyD (Spain) through grants
AYA-0973, AYA-07468-C03-02 and AYA-07468-C03-03
from the PNAyA. We  thank the Centro de Computaci\'on
Cient\'{\i}fica (UAM, Spain) for computing facilities.
AS  thanks  FEDER financial support from UE.

\clearpage

\begin{figure}
\includegraphics[width=\textwidth]{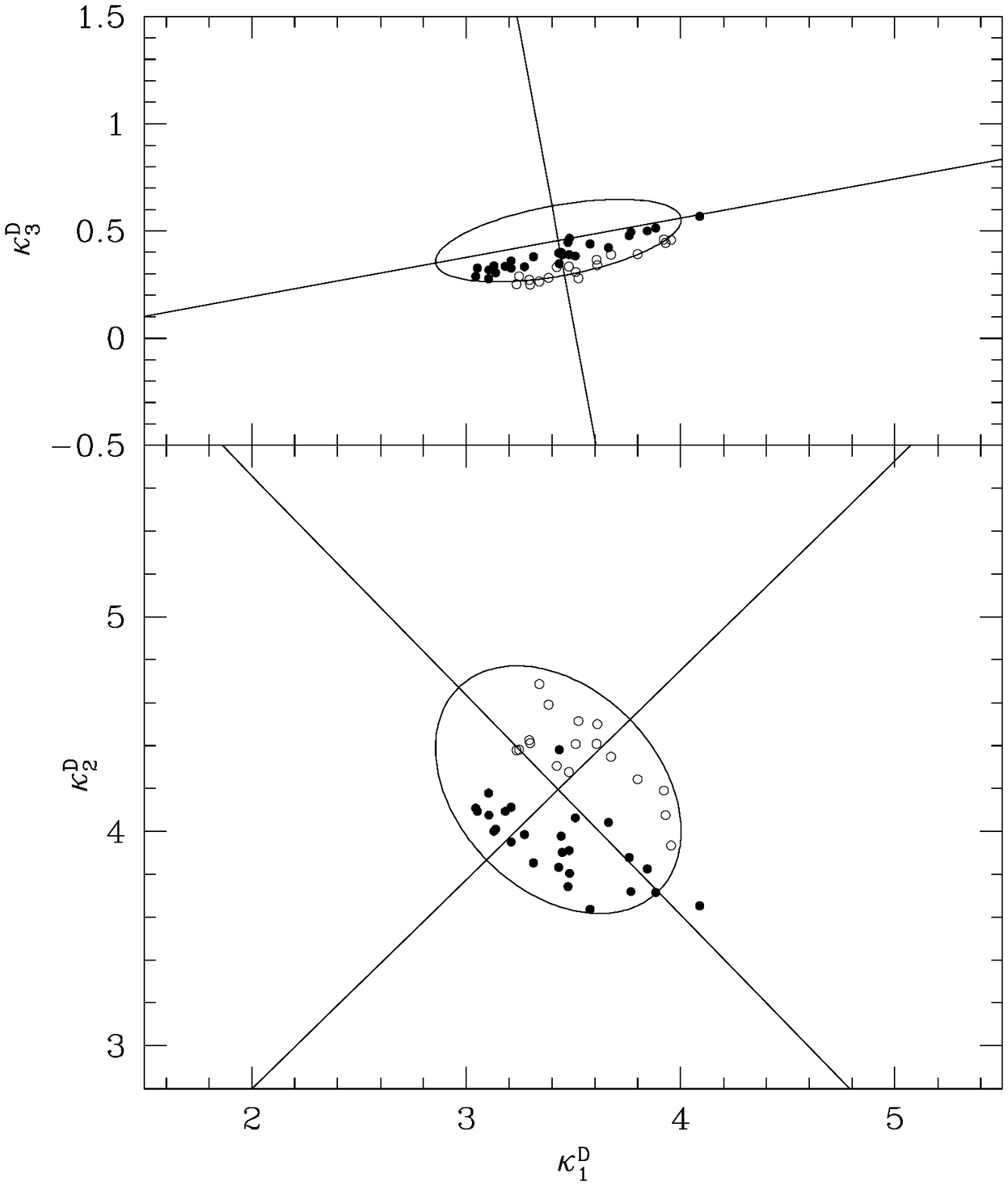}
\caption{Edge-on projection (top panel) and nearly-face-on projection (bottom panel)
of the dynamical FP of ELOs in the 
$\kappa^{D}$ variables (filled circles: SF-A sample;
open circles: SF-B sample).
We also draw the respective concentration ellipses (with their major and minor
axes) for the SDSS early-type galaxy sample from Bernardi et al. (2003b) in the $z$-band.
See text for more details
}
\label{Kplots}
\end{figure}
\newpage

\clearpage

\begin{table}
\caption{}
\begin{center}
\begin{tabular}{lcccc}
\hline
&SF-A &&SF-B  &\\
\hline
\hline
$M_1$               & 0.256 & $\pm$ 0.035& 0.281 & $\pm$ 0.048      \\
$\beta_{\rm vir}$ & 0.221 &   $\pm$ 0.083 & 0.237 & $\pm$ 0.158    \\
$\beta_{\rm M }$  &-0.204 & $\pm$ 0.116 & -0.247& $\pm$ 0.189 \\
\hline
$\beta_{\rm F}$   & 0.025 & $\pm$ 0.048 & 0.022 & $\pm$ 0.081 \\
$\beta_{\rm VD}$  & 0.021 & $\pm$ 0.041 & 0.076 & $\pm$ 0.075 \\
$\beta_{\rm VPC}$ &-0.044 & $\pm$ 0.029 &-0.044 & $\pm$ 0.093 \\
$\beta_{\rm rD}$  &-0.225 & $\pm$ 0.127 &-0.316 & $\pm$ 0.199 \\
$\beta_{\rm rP}$  & 0.019 & $\pm$ 0.009 & 0.016 & $\pm$ 0.017 \\
\hline
\hline

\end{tabular}
\end{center}

%\medskip
         
Column 2: the slopes of the $\kappa_{3}^{\rm D} = M_1 \kappa_{1}^{\rm D} + M_0$
relation (direct fits);
the slopes of the $M_{\rm vir}/M_{\rm bo}^{\rm star}$
and
$c_{\rm i} \propto (M_{\rm bo}^{\rm star})^{\beta_{\rm i}}$
scaling relations for the the SF-A sample,
calculated in $\log - \log$ plots through direct fits.
Column 3: their respective 
95\% confidence intervals. Columns 4 and 5: same as columns 2 and 3
for the SF-B sample.

\end{table}

\end{document}